\documentclass[a4paper]{article}

\usepackage{a4wide}
\usepackage{graphics}
\graphicspath{{img/_build/}}
\usepackage{amsmath}
\usepackage{amsfonts}
\usepackage{amssymb}
\usepackage{listings}
\usepackage{subfig}           
\usepackage[super]{nth}
\usepackage{xfrac}            
\usepackage{xcolor}

\usepackage{authblk}
\usepackage{semantic}
\usepackage{hyperref}

\newcommand*{\definitionname}{Definition}
\newtheorem{definition}{\definitionname}[section]

\providecommand{\keywords}[1]{\noindent\textbf{\textit{Keywords:}} #1}

\newcommand*{\toolname}[1]{\textsc{#1}}
\newcommand*{\highlight}[1]{\colorbox{green!30}{#1}}

\newcommand*{\iatake}{I_Q^{\blacktriangleright}}
\newcommand*{\iaftake}{I_Q^{\blacktriangleright\!\blacktriangleright}}
\newcommand*{\iaro}{I_Q^{\triangleright}}
\newcommand*{\iafro}{I_Q^{\triangleright\triangleright}}

\newcommand*{\expr}[1]{\texttt{\sffamily #1}}  
\newcommand*{\elemname}[1]{\textsf{#1}}        

\newcommand*{\code}[1]{\expr{#1}}

\newcommand*{\valset}{\mathcal{V}}
\newcommand*{\varset}{\mathcal{X}}
\newcommand*{\exprset}{\mathcal{E}}

\DeclareMathOperator{\evalexpr}{eval}
\DeclareMathOperator{\evalae}{Eval}

\newcommand*{\codesize}{\fontsize{8}{7}}
\title{MP net as Abstract Model of Communication for Message-passing Applications}
\author[1]{Martin \v{S}urkovsk\'{y}\thanks{%
This work was supported by The Ministry of Education, Youth and Sports from the National Programme of Sustainability (NPS II) project ``IT4Innovations excellence in science - LQ1602'' and partially supported by the SGC grants No. SP2019/108 ``Extension of HPC platforms for executing scientific pipelines'', and No. SP2018/142 ``Optimization of machine learning algorithms for HPC platform II'', VSB - Technical University of Ostrava, Czech Republic.
  }}
\affil[1]{%
  IT4Innovations\\

  VSB -- Technical University of Ostrava\\
  
  Ostrava, Czech Republic}

\begin {document}

\maketitle

\begin{abstract}
  MP net is a formal model specifically designed for the field of parallel
  applications that use a message passing interface. The main idea is to use MP
  net as a comprehensible way of presenting the actual structure of
  communication within MPI applications. The goal is to provide users with the
  kind of feedback that can help them to check quickly whether or not the actual
  communication within their application corresponds to the intended one. This
  paper introduces MP net that focuses on the communication part of parallel
  applications and emphasizes its spatial character, which is rather hidden in
  sequential (textual) form.
\end{abstract}


\section{Introduction}
MP~net (Message Passing net) is a formal model specifically designed to capture
the communication within MPI\footnote{\url{https://www.mpi-forum.org}} (Message
Passing Interface) applications. MPI is widely used in the field of HPC
(high-performance computing) as a means of communication between nodes of a
computer cluster. Generally speaking, programming parallel application is
significantly harder in comparison to the sequential ones, and the new element
which makes it more difficult is the involved communication. Therefore, a good
understanding of communication is important.

One could ask why care about the communication, the authors know exactly how
they have written the application, don't they? Everyone who has ever written an
application knows how easy it is to make a mistake. The code of MPI applications
actually consists of several parts. Such a program is then distributed among the
nodes of a computer cluster where each node can perform a different part of the
program. In other words, MPI application describes a set of programs that are
spatially distributed, but recorded in a linear form (linear stream of code).
This is one of the aspects which makes programming of distributed applications
more difficult. Moreover, communication involved in such kind of programs gives
rise to a new sort of problems and errors, and programmers have to bear them in
mind. Therefore, it can be very useful to have a tool that provides a different
view of the MPI applications, the view that stresses their spatial character.

The aim is to separate the communication and computational layers of the MPI
code with a focus on the communication part and its presentation via the MP~net.
In this way, to provide the users\footnote{By users we mean the application
  developers.} with a different kind of view of their program that can help them
to confront the actual model of communication with their mental image of it.

\begin{figure}[ht]
  \centering
  \includegraphics[width=.7\linewidth]{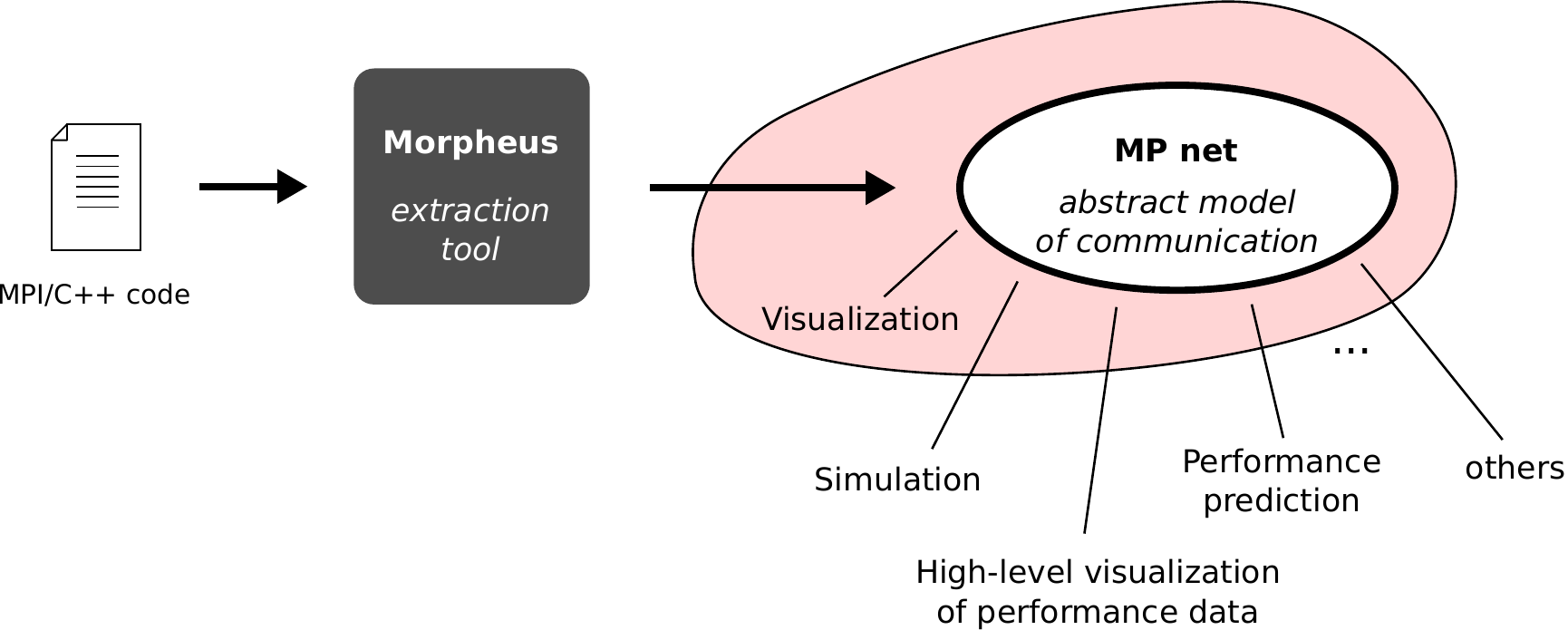}
  \caption{``Big picture'' with the highlighted part considering this paper.}
  \label{fig:main-goal}
\end{figure}

As the MP~net is going to be presented back to users, one of the main
requirements is that the model has to have a ``nice'' visualization. However,
once such a model is available it can be used for other supportive activities
than just the visualization (see \figurename~\ref{fig:main-goal}). Nevertheless,
the first step is to define the abstract model of communication (MP~net) and how
it is extracted from the code of MPI applications.

This paper describes and formalizes MP~nets. The need of a different view of the
communication is demonstrated on a very simple motivational example.
The tool which will implement the presented ideas is currently under
development.

The text is organized as follows. At the beginning, the state of the art is
summarized. Then a motivational example is presented. It shows that a different
kind of view of the communication can be useful even for a very simple program.
The following section introduces and formalizes MP nets. After that we return to
the motivational example and generate MP nets for it. The conclusion sums up the
article and discusses the possible utilization of MP nets.

\section{Related work}

The motivation and original idea for the work presented here comes from the work
on the tool \toolname{Kaira}~\cite{2014-pn,2014-comsis,thesis:bohm}. \toolname{Kaira} is a
development environment for applications using MPI. It strictly
separates the communication and computational layers of the application. The
former one is specified via a visual programming language while the latter one
is written in C++. From the combination of these two inputs the MPI code of the
resulting application is eventually generated. Despite many nice features the
\toolname{Kaira} is provided with, it has not been used by people other than
those from the developer's group. Yet the idea of using a visual form for
describing communication seems to be right because it emphasizes the spatial
character of the communication. However, it turned out that using this visual
form as an input which is provided and created by programmers, is similarly
demanding as writing the application using MPI.

The current work aims to reverse the process and instead of modeling the
communication via a visual language, the goal is to provide an abstract model of
communication for an existing code of an MPI application. In a fully automatic
way without need of any modification in the analyzed program. Hence, it shifts
the paradigm from the field of visual programming to program visualization. In
this way, the programmer is a reader of such model and not its creator.

\begin{figure}[ht]
  \centering
  \includegraphics[width=0.6\linewidth]{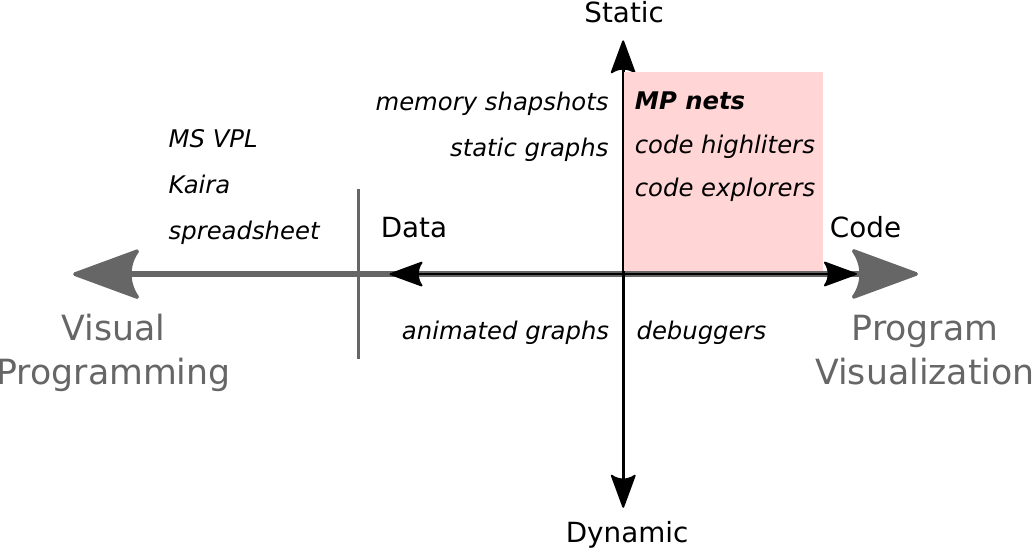}
  \caption{Relationship between visual programming and program visualization with
    taxonomy within program visualization.}
  \label{fig:vp-pv-relationship}
\end{figure}

The genre of visual programming has been defined by Grafton and Ichikawa
\cite{vp1} and includes three distinct areas: graphics techniques that provide
both static and dynamic multidimensional views of software, graphics-based
high-level programming languages, and animation of algorithms and
software~\cite{p11}. Their work was later refined by Myers~\cite{p1}. He
strongly distinguished visual programming, program visualization, and other
high-level programming paradigms. Moreover, he also provided two taxonomies for
the classification of programming systems, one for visual programming and the
other for program visualization. A nice overview of possible program
visualization is summarized in~\cite{Lemieux:2006}. A brief overview of these
categories can be seen in \figurename~\ref{fig:vp-pv-relationship}. Each
category is filled with examples. As can be seen, the \emph{program
  visualization} is divided into four quadrants according to whether the
analysis is done on code or data and whether a static or dynamic aspect is
investigated. MP nets fall into the static-code analysis.

Since this work mainly considers HPC applications, let us look at the supportive
tools in this field. There is a mature category of profiler and trace analyzer
tools. As the representatives of these can be named:
\toolname{TAU}~\cite{TAU:2006}, \toolname{Scalasca}~\cite{scalasca-cube:2007},
\toolname{Vampire}~\footnote{\url{https://vampir.eu/}} and the environment
\toolname{Score-P}~\cite{score-p}. These performance measurement tools can
however be identified as data-static in Myers taxonomy as the analyses are done
during the run time of the application and the result is usually a kind of
graph. As an example of a parallel debugger
\toolname{ARM~DDT}\footnote{\url{https://www.arm.com/products/development-tools/server-and-hpc/forge/ddt}}
can be named.

As for the static-code analyzers whose results are a kind of visualization
presented back to the users, tools such as
\toolname{Sourcetrail}\footnote{\url{https://www.sourcetrail.com/}} or
\toolname{Code~Map}\footnote{\url{https://docs.microsoft.com/en-us/visualstudio/modeling/map-dependencies-across-your-solutions?view=vs-2017}}
can be named. These sort of tools are called \emph{code explorers} and they help
users with orientation in large projects by inspecting the complicated
relationships in the application. The tool that also belongs to this category
and concerns MPI is \toolname{MPI-CHECK}\cite{mpi-check}. \toolname{MPI-CHECK},
however, checks the correct usage of MPI calls.

Most of the currently available tools that provide a view on the communication
are based on space-time graphs~\cite{Lamport:1978}. These tools collect the data
during the run of an analyzed program and provide a visualization based on it.
As far as the author is aware, a tool that would transform the code of an MPI
application into a more comprehensible (visual) form does not exists, yet.

\subsection{Petri nets}
A Petri net is a well-established mathematical modeling language for the
descriptions of distributed systems. A basic form of Petri nets is called
Place/Transition net (P/T net or PTN). A PTN is a 6-tuple $(P, T, I, O, w,
m_0)$, where $P$ is a finite set of \emph{places}, $T$ is a finite set of
\emph{transitions}, $P \cap T = \emptyset$, $I$ is a set of \emph{input arcs} $I
\subseteq P \times T$, $O$ is a set of \emph{output arcs} $O \subseteq T \times
P$, $w$ is a weight function $w: (I \cup O) \to \mathbb{N}$, and $m_0$ is the
\emph{initial marking}; \emph{marking} is a function $m: P \to \mathbb{N}$.

The behavior of such a system is defined through the changes of its state (marking).
Any change of a state in a Petri net is directed by \emph{transition firing rule};
\begin{enumerate}
  \item A transition $t$ is said to be \emph{enabled} at marking $m$ if $\forall p
    \in P: m(p) \geq w(p, t)$.
  \item An enabled transition may fire/occur.
  \item Firing of an enabled transition $t$ changes a marking $m$, written
    $m \xrightarrow{t} m^\prime$, where $m^\prime=m(p) - w(p,t) + w(t,p)$ for all $p \in P$.
\end{enumerate}

\begin{figure}[ht]
  \centering
  \subfloat[][]{%
    \label{fig:ptn-example-a}
    \includegraphics[width=.3\linewidth]{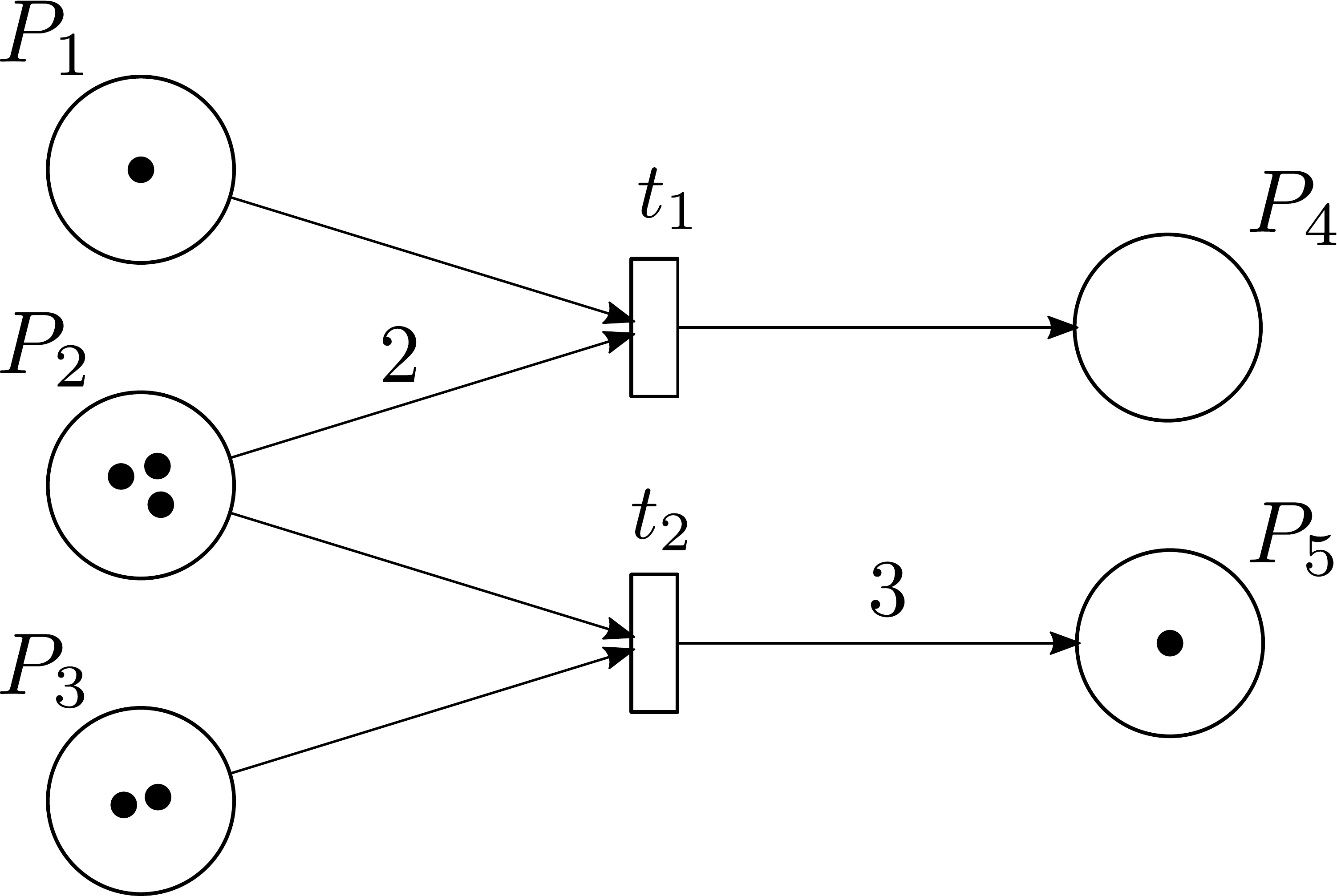}
  }
  \quad
  \subfloat[][]{%
    \label{fig:ptn-example-b}
    \includegraphics[width=.3\linewidth]{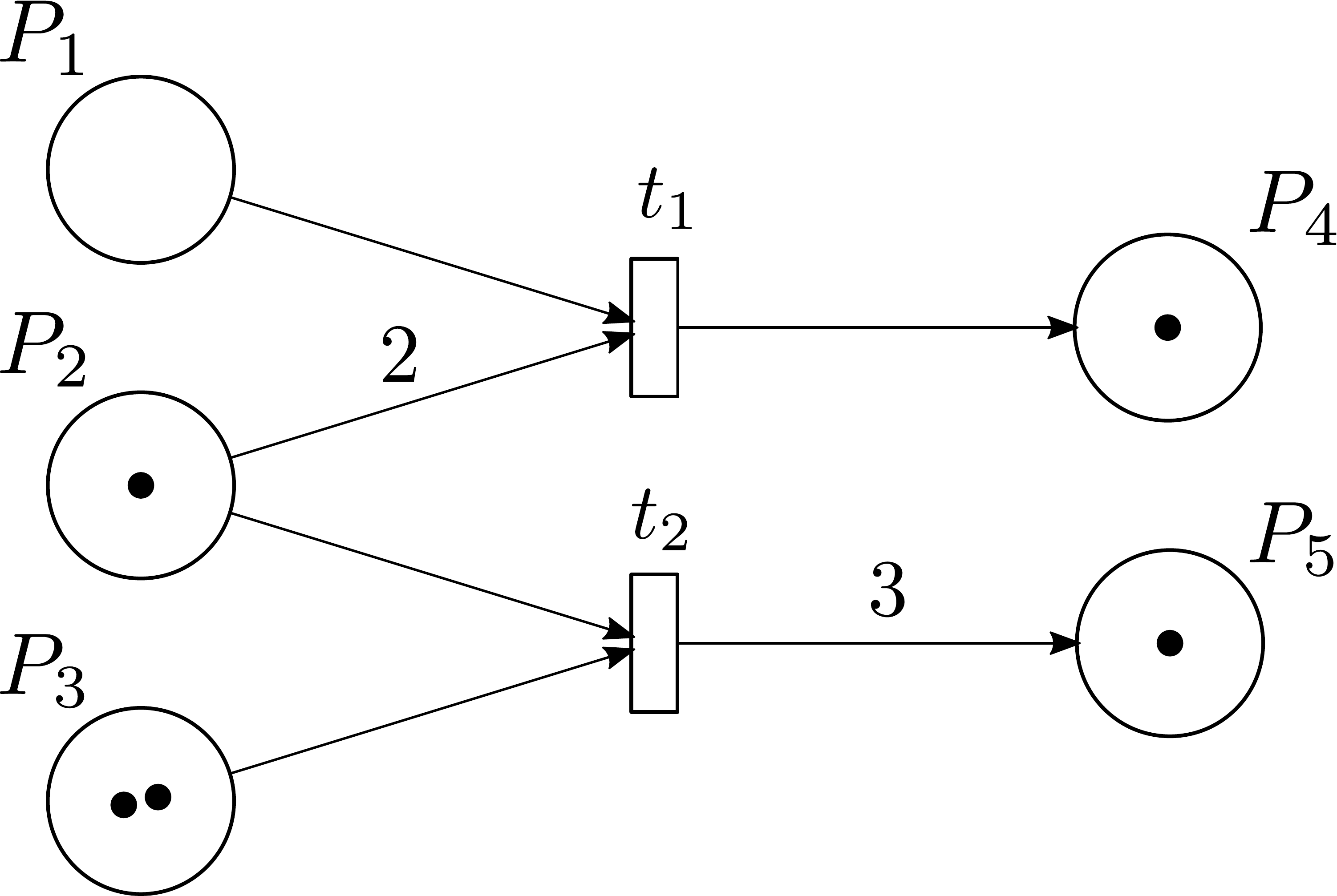}
  }
  \caption[A simple example of PTN system.]{
    \subref{fig:ptn-example-a} A simple example of PTN with an initial marking;
    \subref{fig:ptn-example-b} the change of marking after firing $t_1$.}
  \label{fig:ptn-example}
\end{figure}

One of the advantages of Petri nets is that they have established and
comprehensible visualization. \figurename~\ref{fig:ptn-example} shows an example
of a simple PTN with an initial marking~\subref{fig:ptn-example-a}, and its
change~\subref{fig:ptn-example-b} after firing transition $t_1$.

In a standard way, places are drawn as circular objects while transitions as
rectangular ones. Arcs between places and transitions represent a flow relation,
and numbers at arcs the weight. The weight of an arc without a number is $1$. The
current marking is determined by a distribution of tokens, black dots, among
places.

An extension, called Coloured Petri nets~\cite{book:cpn} (CPNs), enriches the P/T
nets with tokens with an inner structure. The places do not hold just
``black'' opaque tokens, but rather tokens carrying a value. The places in CPNs
then do not contain only a number of tokens but multisets of values. This
involves more complex conditions under which a transition is enabled and what will
happen after its occurrence. Nevertheless, the core idea remains the same;
input places of a transition have to contain an appropriate number of ``right''
tokens. One of the best-known tools based on CPNs is \toolname{CPN
  Tools}\footnote{http://cpntools.org}.

The QPN~\cite{qpn93,qpn2012} (Queuing Petri nets) enriches the concept with
queuing places. In comparison to CPN the data in a place are arranged in a queue
instead of a multiset, hence not just presence of a token in a place is important
but also its position in the queue.

\section{Motivational example}
The main idea and motivation of this work is presented in this section. We will
look at three different implementations of one particular example. For each, we
will look closely at the communication and discuss the differences. It will show
that the first idea we implement might be different from the one we want.
Therefore, a good understanding of communication within a parallel application
is important.

\begin{figure}[ht]
  \centering \includegraphics[width=0.45\linewidth]{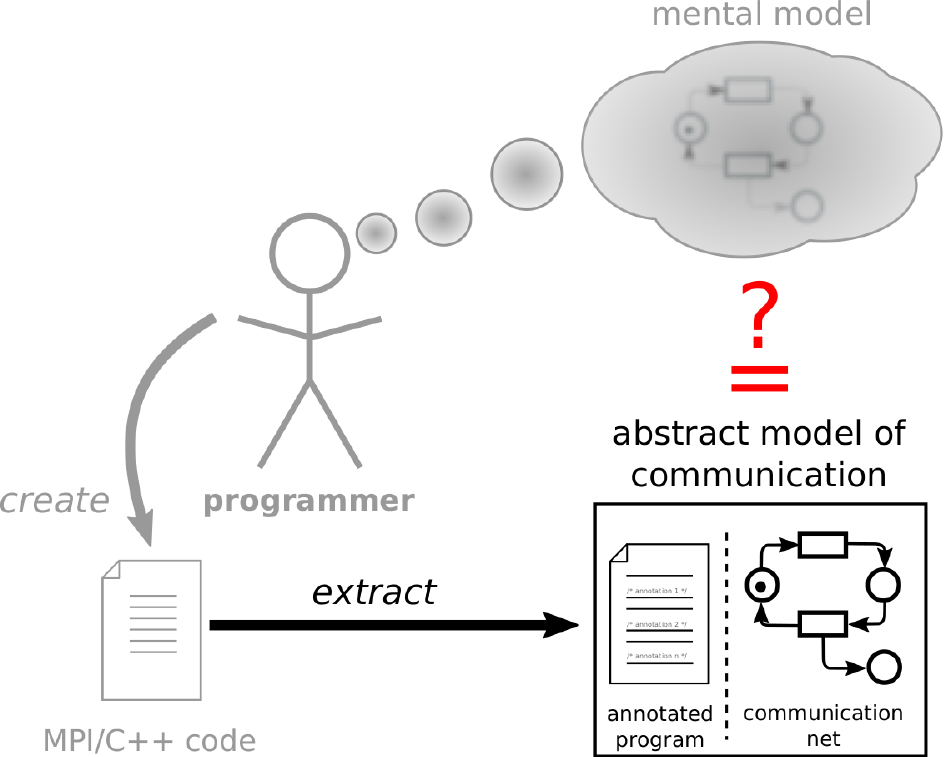}
  \caption{Workflow with extracted abstract models.}
  \label{fig:workflow-with-am}
\end{figure}

Before we start with an example, let us briefly look at
\figurename~\ref{fig:workflow-with-am}. Its gray part (right-hand side) shows a
``standard'' workflow when developing a program. Every programmer has to think
through a problem at the beginning, come up with an idea of what the main
components are, and how they interact with each other or with an external
environment. They have to create a \emph{mental image} of the program that is
going to be developed. The same holds for any programs including parallel ones.

When developing parallel programs programmers have to consider the
communication done during computation. MPI often used to deal with that
communication, is quite low-level. Then it is not surprising that the actual
communication model of the resulting program may differ from the intended one.
The difference between the intended communication model and the actual one --
written in an application's code -- is the main motivation of this work. As can
be seen in \figurename~\ref{fig:workflow-with-am}, the standard workflow is
enriched by a black part -- work with an actual model of communication. At the
moment, the programmer can use that model and confront it with their mental
image.

\subsection {All-send-one example}
\begin{figure}[ht]
  \centering
  \includegraphics[width=.5\linewidth]{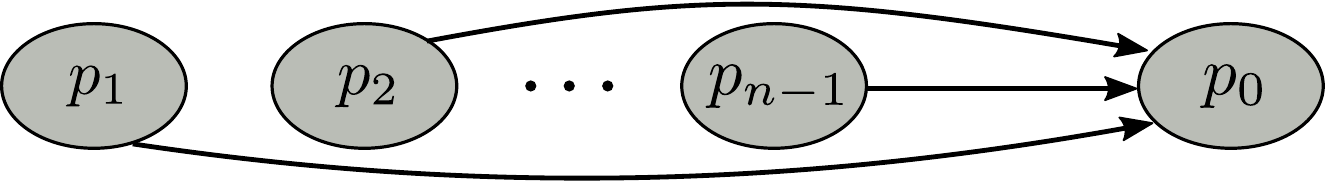}
  \caption{Mental model of the simple task.}
  \label{fig:st-mental-model}
\end{figure}

Let us demonstrate this workflow on a very simple example. Imagine we have a
task to create a parallel program that works on $n$ processes, $p_0, p_1, \dots,
p_{n-1}$. Each process except the $0^{th}$ one sends a message (e.g. its process ID)
to $p_0$ and $p_0$ receives all these messages. Our mental image may look like
the one in \figurename~\ref{fig:st-mental-model}.

With this image in our mind we can make a program
(\lstlistingname~\ref{prg:st-ordered-causal-dependency}) called \emph{all-send-one}, that
may look as follows. We can see, there is a loop on the $0^{th}$ process
($rank=0$) that receives all the messages sent by other processes. But, does the
program behave (communicate) the way we expected? To answer this question, we can
examine the code and state whether it does or does not. Nevertheless, it would
be nice to have a tool that presents communication in the program in a more
comprehensive form than just looking at the source code.

\begin{lstlisting}[
  language=C++,
  caption={First version of \emph{all-send-one} program.},
  label=prg:st-ordered-causal-dependency
]
#include "mpi.h"

int main (int argc, char *argv[]) {
  enum { TAG_A };

  MPI_Init(&argc, &argv);

  int rank, size;
  MPI_Comm_rank(MPI_COMM_WORLD, &rank);
  MPI_Comm_size(MPI_COMM_WORLD, &size);

  if (rank == 0) {
    int ns[size-1];
    for (int i=0, src=1; src < size; i++, src++) {
      MPI_Recv(&ns[i], 1, MPI_INT, src, TAG_A, MPI_COMM_WORLD, MPI_STATUS_IGNORE);
    }
  } else {
    MPI_Send(&rank, 1, MPI_INT, 0, TAG_A, MPI_COMM_WORLD);
  }

  MPI_Finalize();
  return 0;
}
\end{lstlisting}

\begin{figure}[ht]
  \centering
  \includegraphics[width=.5\linewidth]{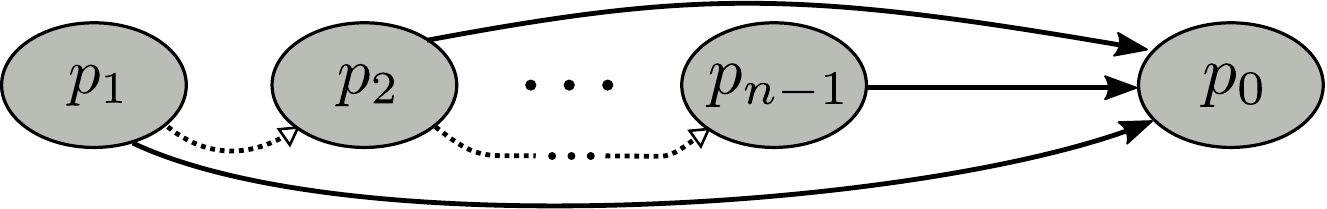}
  \caption{``Extracted'' model of program \ref{prg:st-ordered-causal-dependency}.}
  \label{fig:st-extracted-model-1}
\end{figure}

Let us imagine we have an application that shows us a model of communication as
it really is. By applying it on the code of our \emph{all-to-one} program we
could get a model similar to the one in \figurename~\ref{fig:st-extracted-model-1}.
It is visible that there is a difference between the extracted model and our
mental image; causal dependencies between individual sends. This is caused by
the receiving loop (lines 14-16), because all the messages are received
gradually in one specific order. At this moment, the programmer can see the
difference and it is their decision whether to make some changes in the original
program or not.

Suppose we want to get rid of these causal dependencies. The very first thing
we can notice is that the receiving operation waits for a specific source
(argument \code{src} on line 15). Actually, we don't need to wait for a specific
source. Instead, we can use the \code{MPI\_ANY\_SOURCE} wild card; the
process which sends the data as the first one, is the one which the data is
received from. The modification can be seen in the following program
(\lstlistingname~\ref{prg:st-unordered-causal-dependency}).

\begin{lstlisting}[
  language=C++,
  caption={\emph{all-send-one} using \code{MPI\_ANY\_SOURCE} wild card.},
  label=prg:st-unordered-causal-dependency,
  firstnumber=13, firstline=13, lastline=16
]
    int ns[size-1];
    for (int i = 0; i < size - 1; i++) { // don't need "src" variable
      MPI_Recv(&ns[i], 1, MPI_INT, (*\highlight{MPI\_ANY\_SOURCE}*), TAG_A, MPI_COMM_WORLD, MPI_STATUS_IGNORE);
    }
\end{lstlisting}

\begin{figure}[ht]
  \centering
  \includegraphics[width=.5\linewidth]{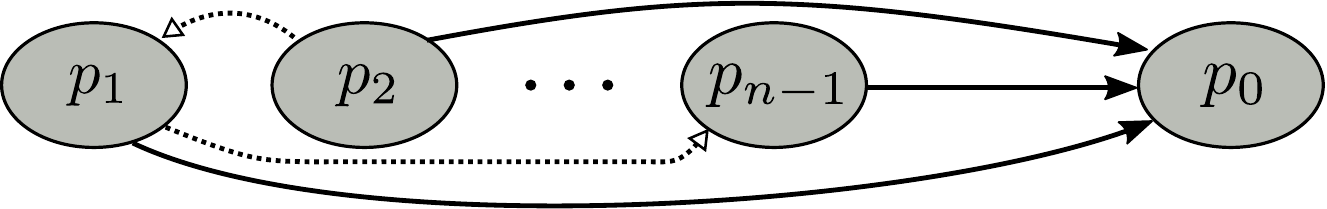}
  \caption{``Extracted'' model of program \ref{prg:st-unordered-causal-dependency}.}
  \label{fig:st-extracted-model-2}
\end{figure}

The question is whether the change we have made solved the problem with the
causal dependency. The answer is ``yes and no''. Look at
\figurename~\ref{fig:st-extracted-model-2}. Although the specific order of
receiving messages was removed, each receive is completed within the cycle,
hence before the next receive can start, the previous receive request has to
complete. Therefore, the data is still received gradually (one by one). The only
change is that the data is received in a non-deterministic order.

Another change we can do is to send the receive requests in a non-blocking way
and then wait for all of them (\lstlistingname~\ref{prg:st-broken-causal-dependency}).In
this way we still have to wait for all the messages, but the way how the
messages are received is left on the used implementation of MPI. So, we can
say, that the last version is the closest one to the original idea
(\figurename~\ref{fig:st-mental-model}).

\begin{lstlisting}[
  language=C++,
  caption={\emph{all-send-one} using non-blocking communication to break causal
    dependency.},
  label=prg:st-broken-causal-dependency,
  firstnumber=13, firstline=13, lastline=18
]
    int ns[size-1];
    MPI_Request requests[size-1];
    for (int i=0, src=1; src < size; i++, src++) {
      (*\highlight{MPI\_Irecv}*)(&ns[i], 1, MPI_INT, src, TAG_A, MPI_COMM_WORLD, &requests[i]);
    }
    (*\highlight{MPI\_Waitall}*)(size-1, requests, MPI_STATUSES_IGNORE);
\end{lstlisting}

As we have seen, even in a very simple example the actual model of communication
may differ from the intended one. In this particular example, the first naive
implementation can be considered as a mistake. Nevertheless, it would not cause
any failure. On the other hand, such mistakes can be seen as even worse in
comparison to those that cause an application crash, because it can take a
significant amount of time when they begin to show up. The main goal of this
work is to provide a tool that helps programmers keep track of what the
actual communication looks like in their programs and not to fall into a state
where they think ``the communication is alright''.

\section{MP net}
MP net, proposed in this text, is a formal model specifically designed for the
field of parallel applications that use MPI. The purpose of MP net is to provide
a different view of MPI applications that is focused rather on their
communication part than the computational one. In other words, it exposes the
spatial character of the communication compared to the linear record of standard
programming languages like C/C++.

MP net is built on Coloured Petri nets~\cite{book:cpn} (CPNs) and a
simplified version\footnote{The stochastic and timed aspect of QPN is discarded.
  From the perspective of MP net, only scheduling of tokens in queuing places is
  important.} of Queuing Petri nets~\cite{qpn93,qpn2012} (QPNs). The goal
is to propose a model that is, on one hand, powerful to fully capture the
semantics\footnote{In the current version the semantics of point-to-point
  communication is mainly considered.} of MPI~\cite{MPIStandard}, and, on
the other hand, simple enough to be understood relatively easily by people. The
balance between these two main requirements is the reason why a new model is
proposed. However, there is no reason ``to re-invent the wheel''; therefore,
MP~net is defined through the specification of its structure (syntax) and the
behavior (semantics) is defined by a set of transformations into the simplified
version of QPN.

\subsection{Basic definitions}
For any set $A$, $A^\ast$ is a free monoid over this set with the operation of
\emph{sequence concatenation}: $\cdot: A^\ast \times A^\ast \to A^\ast$ defined
as $\forall s, s^\prime \in A^\ast: s \cdot s^\prime = s^{\prime\prime}$ where
$s^{\prime\prime} \in A^\ast$, and zero element \emph{empty sequence}: $\epsilon
\in A^\ast$. For any sequence $s=[s_1,s_2,\dots,s_n] \in A^\ast$, $s_i$ is the
\emph{$i^{th}$ element} of the sequence where $i\in \{x\in \mathbb{N}\ |\ 1 \leq
x \leq n\} = \mathbb{N}_n$, and $|s|$ is the \emph{length} of the sequence. For
any $op: A -> B$ and a sequence $s=[s_1,s_2,\dots,s_n] \in A^\ast$, let us
define $op(s):=[op(s_1), op(s_2), \dots, op(s_n)] \in B^\ast$.
An \emph{application of a function}, $f: A \times \prod_{P\in \mathcal{P}}P \to
B$, \emph{on a sequences} $s=[s_1, s_2, \dots, s_n] \in A^\ast$ is defined as
$f(s) := [f(s_1,p_1,\dots,p_{m}), f(s_2,p_1,\dots,p_{m}), \dots,\allowbreak
f(s_n,p_1,\dots, p_{m})] \in B^\ast$ where $m=|\mathcal{P}|$.

\subsection{Structure of MP net}\label{mpn-structure}

\newcommand*{\qu}{\mathrm{QU}}
\newcommand*{\qst}{\mathrm{ST}}
\newcommand*{\qtr}{\mathrm{TR}}
\newcommand*{\qac}{\mathrm{AC}}
\newcommand*{\qtn}{\mathrm{TN}}

\newcommand*{\qcn}{\mathrm{DM}} 

\begin{definition} \label{def:mpnet}
  \emph{MP net (MPN)} is a finite set of addressable areas, where each area is a
  pair of an \emph{annotated sequential program fragment} and \emph{communication net}: \[
  \mathrm{MPN} = \left\{(\mathcal{P}_1, \mathrm{CN}_1), (\mathcal{P}_2,
    \mathrm{CN}_2), \dots, (\mathcal{P}_n, \mathrm{CN}_n) \right\}
  \]
\end{definition}

An annotated sequential program fragment arises from an input program code. The program is
split into the fragments according to conditions directed by \code{rank}
value\footnote{The value provided by \code{MPI\_Comm\_rank} call.}. In fact, the
fragments represent subprograms performed on particular computational units
(ranks). The segments transformed into a CN are highlighted. There are one or
more elements within the CN for boldfaced expressions. All such expressions are
followed by an annotation. The annotation comprises a \emph{label} and
\emph{sequence of directives}. Labels help with orientation in CN. All the
elements that appear in a net because of a particular annotated expression use
the label as a prefix for their name. A sequence of directives set up data exchange
between the program and communication net. There are three allowed directives
with the following syntax:
\begin{itemize}
  \item \code{put(place=expression)}
  \item \code{wait(place)}
  \item \code{get(place $\to$ variable)}
\end{itemize}.

Communication net is built upon a simplified version of Queuing Petri
net~\cite{qpn93,qpn2012}, defined in \definitionname~\ref{def:sqpn}.
Transformation into the simplified version of QPN is presented in
Section~\ref{mpn-behavior}

\begin{definition}
A \emph{Communication Net (CN)} is a triple $\mathrm{CN} = (\mathrm{CPN}, P_Q,
I_Q, A_{\mathrm{CF}})$, where

\begin{itemize}
  \item $\mathrm{CPN} = (P, T, C, I, O, M_0)$ is the underlying Coloured Petri net.
  \item $P_Q \subseteq P$ is a set of queuing places.
  \item $I_Q \subseteq P_Q \times T$ is a set of queuing-input arcs. The set is
    composed of four distinguished types: $I_Q = \iatake \cup \iaftake
    \cup \iaro \cup \iafro$ where $\iatake \cap \iaftake \cap \iaro \cap \iafro
    = \emptyset$.
  \item $A_{\mathrm{CF}} \subseteq X \times Y$ a set of control flow arcs where
    $X,Y \in \{P, T\}$.
\end{itemize}
\end{definition}

The graphical representation is similar to the one of CPNs. Unlike
QPNs\footnote{QPNs depict queuing places as circles with a horizontal line
  splitting it into two parts which reminds us that the queuing place is
  composed of two parts: the queue and depository.}, queuing places are depicted
in CNs as rounded objects with a white background, i.e., as standard places in
CPNs. The multiset places are then distinguished by gray color.

There are several types of arcs. The first category forms input arcs leading
from queuing places, $I_Q$. This category is formed by four different types. The
types are distinguished by a used arrow head which corresponds to upper indexes
of particular disjoint sets. A combination of a queuing place and a particular
type of input arc determines the strategy used to access data in the queue.
Conditional arcs, i.e., arcs with an expression preceded by a conditional one
are emphasized using dashed style. The last type of arcs represents control flow
arcs, $A_{\mathrm{CF}}$. These help a user to orient in the communication net
while the control is given back to the program. In fact, the control flow arcs
just have an informative character without any effect and are removed during the
transformation\footnote{Therefore, control flow arcs can connect the
  same net elements.}. This type of arcs is depicted as tiny gray arrows.

\subsection{Arc expressions}
Let us define a set of values $\valset$ and a set of variables $\varset$. There
are three types of values that are always part of the value set; \expr{unit} for
a Unit value, \expr{true} and \expr{false} for Boolean values, and natural
numbers $\mathbb{N}$, i.e., $\{\expr{unit}, \expr{true}, \expr{false}\} \cup
\mathbb{N} \subseteq \valset$. These values and objects composed of them, e.g.,
tuples or records are considered to be \emph{transparent}. The structure of
transparent object can be inspected. There is another category of values, let us
call them \emph{opaque}. Opaque values arise from the evaluation of expressions
used in a source language, e.g., C++. They are viewed as binary objects without
any inner structure.

A set of all expressions used within MP net as $\exprset$. Values and variables
are considered to be expressions too, $\valset, \varset \subseteq \exprset$. To
evaluate an expression to the value a function $\evalexpr: (\exprset \times
\varset -> \valset) -> \valset$ is defined. It takes an expression $e \in
\exprset$ and binding $b: \varset -> \valset$ that specifies a binding function
assigning values to free variables of the expression. A set of free variables of
an expression is obtained by the function: $FV: \exprset -> \varset$. To
evaluate the entire arc expression, the function $\evalae: (2^\exprset
\times \exprset^\ast) \times (\varset -> \valset) -> \valset^\ast$ is defined. \[
    \evalae((C, \tau), b) = \begin{cases}
      \epsilon & \mbox{if}\ \exists\,c\in C: \evalexpr(c,b) = \expr{false} \\
      \evalexpr(\tau,b) & \mbox{otherwise} \\
  \end{cases}
\]

\paragraph{Input arc expressions} are assigned via the function $ie$, $ie: I ->
(2^\exprset \times (\exprset^\ast \times (\varset \cup \mathbb{N}) \times
\varset))$. It composes of two parts: a \emph{set of conditions} and \emph{main
  expression}. The used syntax is the following:

\begin{center}
\code{[conditions] (pattern, size, values)}
\end{center}

\begin{enumerate}
  \item \code{conditions} -- a set of conditions $C \in 2^\exprset$.

  \item \code{pattern} -- a sequence of expressions $[e_1, e_2,\dots,
    e_n] \in \exprset^\ast$ representing value(s) of token(s) that is
    taken from the corresponding place.
  \item \code{size} -- an expression specifying of how many tokens is
    desired. It can be a natural number or free variable. 
  \item \code{values} -- a free variable for actual values.
\end{enumerate}

Let us define two derived forms of the input-arc's sub-expression, so that:
$(pattern, size) := (pattern, size, \_)$ and $pattern := (pattern,\allowbreak
|pattern|, \_)$. The former one allows the values variable to be omitted
while the latter one omits also the size, in this case the size
of the pattern is used.

\paragraph{Output arc expressions} are assigned via the function $oe$, $oe: O ->
(2^\exprset \times \exprset^\ast \times A)$. It composes of three parts: a
\emph{set of conditions}, \emph{main expression}, and \emph{location}
(addressable area). The syntax of the arc expression is:

\begin{center}
  \code{[conditions] pattern@location}
\end{center}

\begin{enumerate}
  \item \code{conditions} -- a set of conditions $C \in 2^\exprset$.

  \item \code{pattern} -- a sequence of expressions $[e_1, e_2,\dots , e_n]
    \in \exprset^\ast$ representing value(s) of token(s) that are put into the
    connected place.

  \item \code{@location} -- addressable are $a \in \mathrm{MPN}$. It may be
    omitted, but in this case a default addressable is used, i.e.,
    addressable area of the arc's corresponding place.
\end{enumerate}

\subsection{Behavior of MP net}\label{mpn-behavior}
Semantics of MP net is given by a set of transformations into a simplified
version of Queuing Petri net. There are three parts: a program, a communication
net, and addressable areas whose transformation will be solved separately.

\subsubsection{Transformation of annotated sequential program}
Any program can be represented in a form of state machine where each node
represents the current state of the program and directed edges the statements
that change the state. The state is composed of a position in such a graph and
the current state of the memory. Such a graph can be transformed into a Coloured
Petri net straightforwardly. Each node in the original graph is replaced with a
unit place and all directed edges with transitions connected to corresponding
input/output places. For further reference, let us call these arcs \emph{control
  arcs}. To capture the memory state, one colored place is added. Each
transition is then connected with this ``memory place'' via an input and output
arc. An input arc takes a current state of the memory and returns a modified
version according to a program statement. Let us call such a resulting CPN,
\emph{program net}.

The directives that extend the original program are transformed in the following
way. For each directive a new series of unit place and transition is always
added. With this series the output control arc of the transition representing
the annotated statement is replaced. If there are more directives in an
annotation, the output control arc of the transition representing the preceding
directive is replaced. The newly added transition is connected to the rest of
the MP net as follows:

\begin{itemize}
\item \code{put(place=expression)} -- connects the transition with the memory
  place so that it takes the actual value and without any modification is
  returned back. Moreover, from the transition a new arc leads to the place
  specified by the put directive within the communication net, using the
  specified expression.
\item \code{wait(place)} -- adds a new read-only input arc to the transition
  leading from the specified place. The transition then synchronizes the program
  net with the communication one but does not modify the content of the place.
\item \code{get(place $\to$ variable)} -- connects the transition with the
  memory place in the similar way as with the put directive, but this time a new
  input arc from the specified place is added. This arc is a bulk arc which
  takes the entire content (possibly empty) of the place whenever the transition
  occurs. Therefore, the value of specified variable is set by the content of a
  specified place.
\end{itemize}

\subsubsection{Transformation of the communication net}

A communication net is transformed into a simplified version of Queuing Petri net.
Therefore, before we start, let us rephrase the definition of Queuing Petri
net and queue from~\cite{qpn93}. The definition is simplified in comparison
to the original one as it is considered neither a timed nor stochastic aspect of
QPN. For the purpose of MP net, queuing places and a way of scheduling tokens in
the queue need to be established.

\begin{definition}\label{def:queue}
$\qu(p)$ defines a \emph{queue} as a 4-tuple $\qu(p) = (\qst(p), \qtr(p),\allowbreak
\qac(p), \qtn(p))$ where

\begin{itemize}
  \item $\qst(p)$ specifies the feasible states of a queue.
  \item $\qtr(p): (\qst(p) \times \qcn(p)) \times (\qst(p) \times \qcn(p))
    \times C(p) \to 2$. This two-state value can be interpreted as decision
    procedure whether or not the token is serviced.
  \item $\qac(p): \qst(p) \times C(p) \to \qst(p)$ specifying the next state of
    the queue after arrival of a token of color $\in C(p)$. 
  \item $\qtn(p): \qst(p) \times C(p) \to \mathbb{N}_0$ specifying the number of
    tokens in the queue.
\end{itemize}
\end{definition}

\begin{definition}\label{def:sqpn}
\emph{Simplified Queuing Petri net} is defined as a pair $\mathrm{SQPN} = (\mathrm{CPN}, Q)$
where

\begin{itemize}
  \item $\mathrm{CPN}$ is the underlying Coloured Petri net,
  \item $Q=(\bar{Q}, (q_1,\dots, q_{|P|}))$ where
  \begin{itemize}
    \item $\bar{Q} \subseteq P$ is the set queuing places,
    \item $q_i =
        \begin{cases}
          \qu(p_i) & \text{if } p_i \in \bar{Q} \\
          \bot     & \text{if } p_i \in P \setminus \bar{Q}
        \end{cases}$
  \end{itemize}
\end{itemize}
\end{definition}

By defining $q_i = \bot$ for $p_i \in P \setminus \bar{Q}$ is meant that there
is no scheduling strategy for this kind of places. In other words, arriving
tokens are put directly into the depository. In QPN each place is composed of a
queue and depository.

Tokens located in a queue are not accessible for the net. In order to make them
accessible, tokens need to get into the depository from the queue. $\qcn(p)$
specifies the current marking of the depository. $\qcn(p) := [C(p) \to
\mathbb{N}_0]$ is the set of all functions $f: C(p) \to \mathbb{N}_0$. The
action of moving a token from a queue to the depository is called service of a
token and it is specified via the $\qtr(p)$ function.

Besides the transition enableness and occurrence which determine the dynamic
aspect of petri-net like systems, QPN defines a phase of \emph{scheduling in
  queuing places}. During this phase, the tokens that are ready to serve are
moved from the queue to the depository. The check whether or not a specific
token is ready is done via $\qtr(p)$ function, $p \in P$. All tokens for which
$TR(p)((s, d), (s^\prime, d^\prime), v) > 0$ can be served, where $(s,d)$ is the
state of the place's queue and depository and $(s^\prime, d^\prime)$ is the
state of the place's queue and depository after serving token $v$.

\paragraph{Communication net to SQPN} -- Queuing places of CN
corresponds to the queuing places of SQPN, $\bar{Q} = P_Q$. The definition of
queue types ($q_i$), i.e., definitions of feasible states ($\qst$), value
arriving function ($\qac$), and token number function ($\qtn$) for queue places
look as follows: $\forall q \in P_Q:$
\begin{itemize}
  \item $\qst(q) = C(q)^\ast$
  \item $\qac(q)(s,v) = s{\cdot}v$ (concatenation)
  \item $\qtn(q)(s,v) = \begin{cases}
      0                   & \mathrm{if\ } |s| = 0 \\
      1 + \qtn(q)(s_1, v) & \mathrm{if\ } s = v{\cdot}s_1 \\
      \qtn(q)(s_1, v)     & \mathrm{if\ } s = v^\prime{\cdot}s_1, v \neq v^\prime
      \end{cases}$
\end{itemize}
The definition of $\qtr$ depends on the type of queuing input arc. There are four
distinguished types: $\iatake$, $\iaftake$, $\iaro$, and $\iafro$. These can be
categorized into two groups: \emph{single-headed} ($\iatake \cup \iaro$) and
\emph{double-headed} ($\iaftake \cup \iafro$) arcs.

\begin{figure}[ht]
  \centering
  \subfloat[][]{%
    \label{fig:single-headed-case}
    \includegraphics[width=0.4\linewidth]{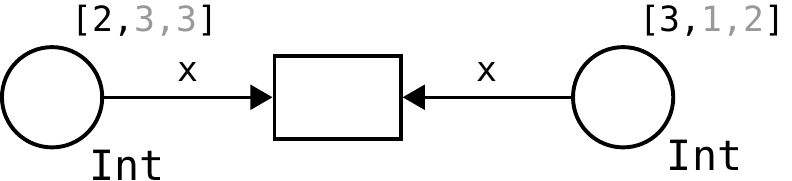}
  }
  \quad
  \subfloat[][]{%
    \label{fig:double-headed-case}
    \includegraphics[width=0.4\linewidth]{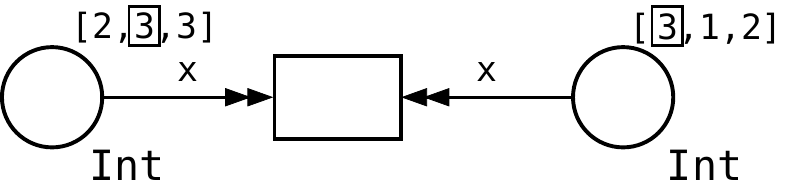}
  }
  \caption[Single- vs. double-headed arcs.]{
    \subref{fig:single-headed-case} Single-headed arcs match only values on the
    head of queue (marked by angle bracket);
    \subref{fig:double-headed-case} double-headed arcs can ``look further'' in
    the queue (the transition is enabled with binding: $x \mapsto 3$).
  }
  \label{fig:single-vs-double-headed-arcs}
\end{figure}

\begin{itemize} \setlength{\itemsep}{1.2em}
  \item $\forall (q, t) \in \iatake \cup \iaro:
    \qtr(q)((s, d), (s^\prime, d^\prime), v) =~\\
    \begin{cases}
      1 & \mathrm{if\ } \parbox[t]{.8\textwidth}{
      $d^\prime = d + \delta_v, d = \mathbf{0}$ and\\
      $\exists v \in C(q), s_1 \in C(q)^\ast: s = v{\cdot}s_1$ and $s^\prime = s_1$
      }\\
      0 & \mathrm{otherwise}
    \end{cases}$

  \item $\forall (q, t) \in \iaftake \cup \iafro:
    \qtr(q)((s, d), (s^\prime, d^\prime), v) =~\\
    \begin{cases}
      1 & \mathrm{if\ } \parbox[t]{.8\textwidth}{
        $d^\prime = d + \delta_v, d(v) = 0$ and\\
        $\exists v \in C(q), s_1, s_2, \in C(q)^\ast: s = s_1{\cdot}v{\cdot}s_2$ and $s^\prime = s_1{\cdot}s_2$,\\
        s.t. $\nexists s_{11},s_{12} \in C(p)^\ast: s_1 = s_{11}{\cdot}v{\cdot}s_{12}$
      }\\
      0 & \mathrm{otherwise}
    \end{cases}$
\end{itemize}
where $\delta_v$ denotes the Kronecker function.
\figurename~\ref{fig:single-vs-double-headed-arcs} shows a simple example
demonstrating the queue behaviour using single-headed and doubles-headed arcs.

The empty-headed arcs ($\iaro$ and $\iafro$) represent read-only arcs. These
arcs do not modify the content of the depository place. In other words, there is
a back arc returning the same value to the depository.

\subsubsection{Transformation of the MP net}
\begin{figure}[ht]
  \centering
  \subfloat[][]{%
    \label{fig:compound-place}
    \includegraphics[width=0.35\linewidth]{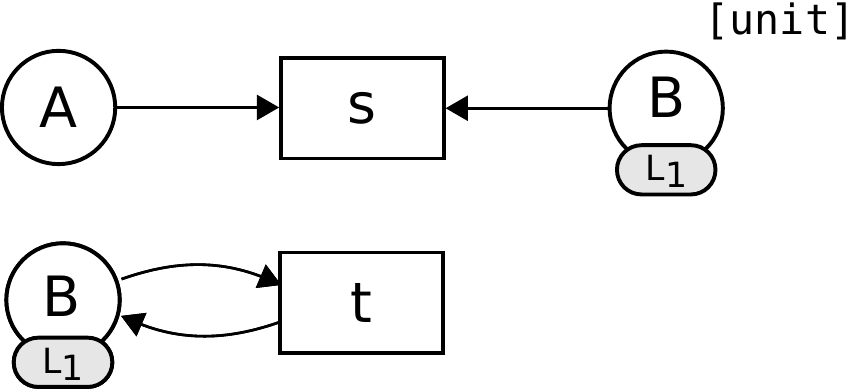}
  }
  \quad
  \subfloat[][]{%
    \label{fig:merged-cp}
    \includegraphics[width=0.35\linewidth]{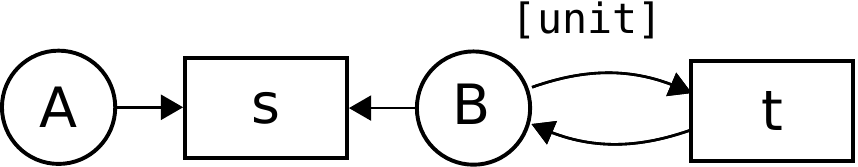}
  }
  \caption[Transformation of compound places.]{
    \subref{fig:compound-place} A simple net using two compound places with a label~$L_1$;
    \subref{fig:merged-cp} The same net without compound places.
  }
  \label{fig:cp-transformation}
\end{figure}

MP net is defined in \definitionname~\ref{def:mpnet} as a finite set of
addressable areas. At the first sight, these areas represent separate nets.
Nevertheless, addressable areas are interconnected via either the output arcs
with a location\footnote{A different location} or compound places. Compound
places represent shared memory in the entire system; therefore, places with the
same compound place label refer to the one and the same place no matter of its
location. This situation is illustrated in
\figurename~\ref{fig:cp-transformation}. The arcs with a specified location do
not put the token(s) into the directly connected output place, but into the
corresponding place into the particular addressable area instead.

\begin{figure}[!ht]
  \centering
  \subfloat[][]{%
    \label{fig:location-arc}
    \includegraphics[width=0.4\linewidth]{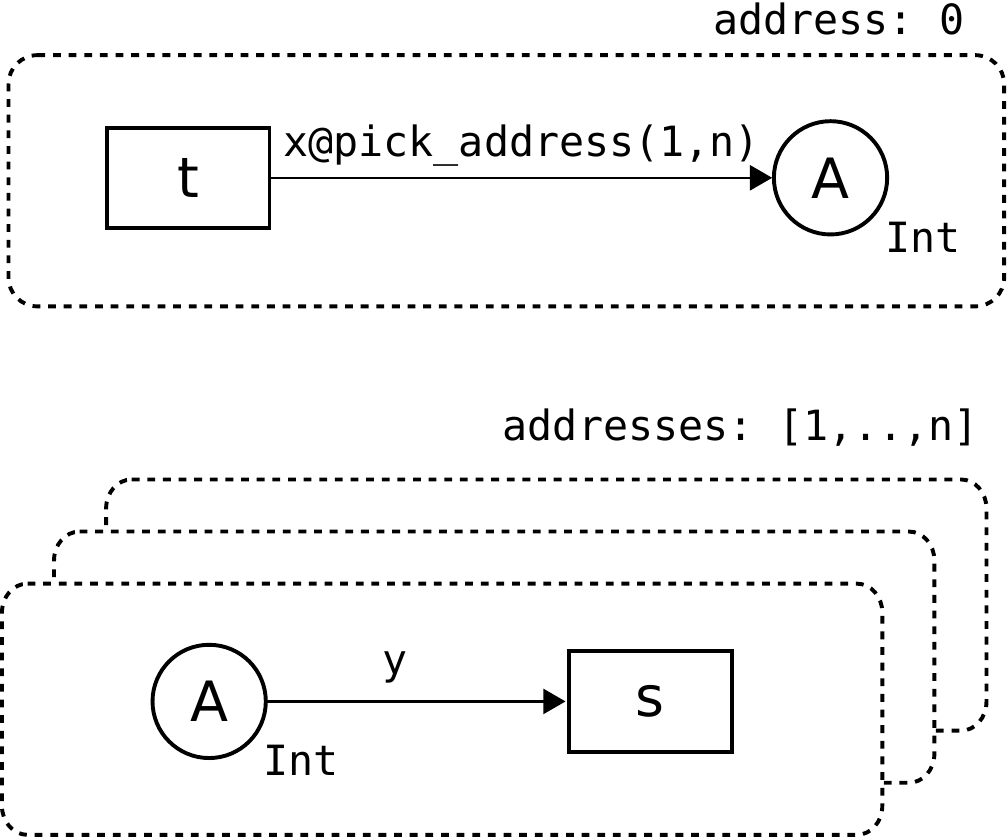}
  }
  \quad
  \subfloat[][]{%
    \label{fig:arc-without-loc}
    \includegraphics[width=0.45\linewidth]{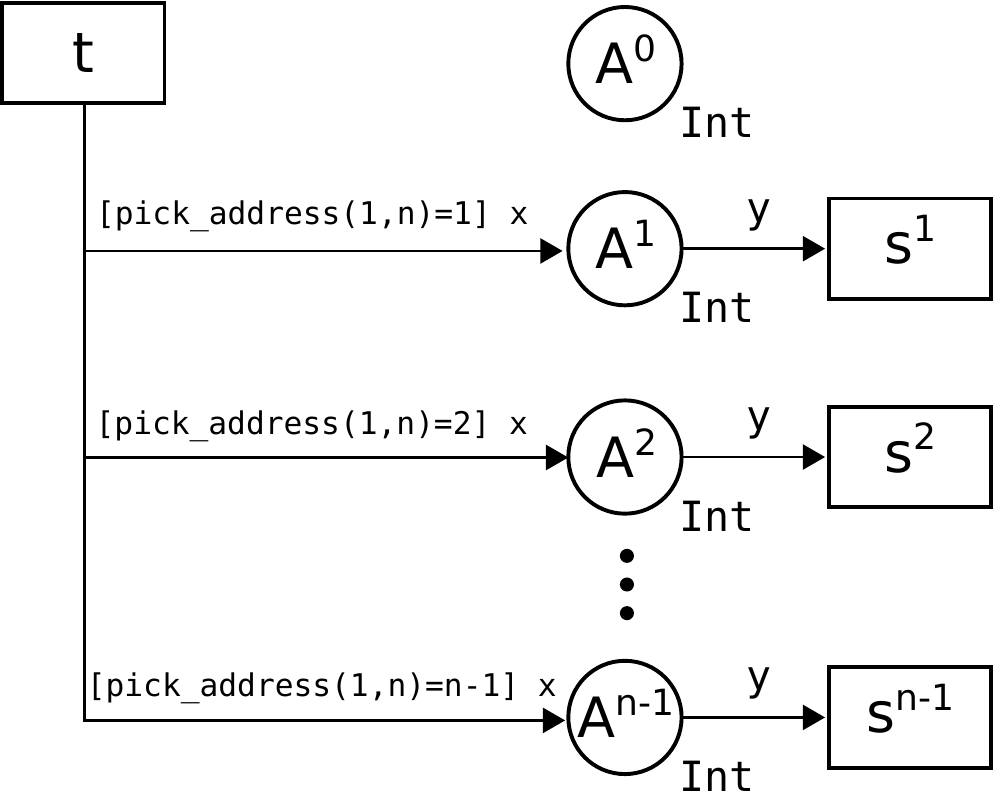}
  }
  \caption[Transformation of location arcs.]{
    \subref{fig:location-arc} A simple MP net using arc with specified location;
    \subref{fig:arc-without-loc} The same net after transformation of
    the arc with specified location.
  }
  \label{fig:loc-arc-transformation}
\end{figure}

Generally, the arcs with a specified location are converted as follows. For each
arc using \expr{`@'}-sign there is a conditional arc leading from the transition
to all corresponding places in different addressable areas. The expression
behind the \expr{`@'} is moved to the condition and checked against the address
of a particular output place. One big SQPN capturing the
behavior of the MP net is composed in this way. This situation is depicted in
\figurename~\ref{fig:loc-arc-transformation}. The used function
\expr{pic\_address} pick an address from the specified range. The upper indexes
of element names (\figurename~\ref{fig:arc-without-loc}) refer to the original
addresses.

\section{Motivational example in MP nets}

\begin{figure}[ht]
  \centering
  \includegraphics[width=.5\linewidth]{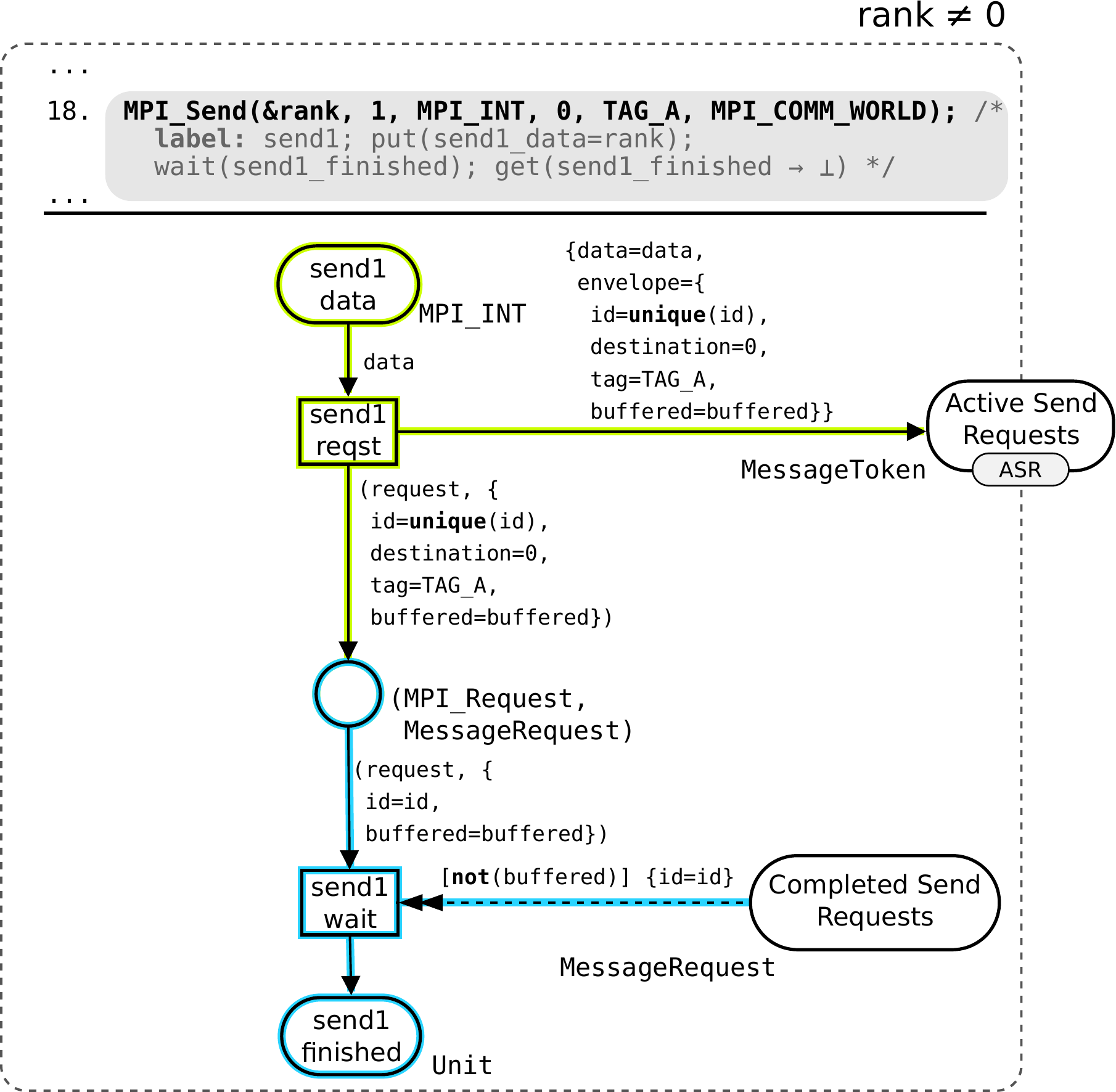}
  \caption{MP net for all-to-one program -- $rank \neq 0$.}
  \label{fig:motiv-example-stage1-rank-gt-zero}.
\end{figure}

Let us look at the MP net for all-to-one program. It consists of two parts. The
first one (see \figurename~\ref{fig:motiv-example-stage1-rank-gt-zero})
represents all the processes that send its rank to process zero. It can be seen,
that the \code{MPI\_Send} call actually consists of two parts. Sending the
request and waiting for its completion (green and blue part of the picture).
There are two special places: \elemname{Active Receive Requests} and
\elemname{Completed Receive Requests}. The former one represents a queue of
requests that is visible for the MPI message broker. It is labeled by compound
place label \elemname{ASR}, i.e., all such places form different views of the
same place. The latter place then represents a local place where the message
broker sends the information about completion of the request.

\begin{figure}[!ht]
  \centering
  \subfloat[][]{%
    \label{fig:motiv-example-stage1-rank-zero}
    \includegraphics[width=.45\linewidth]{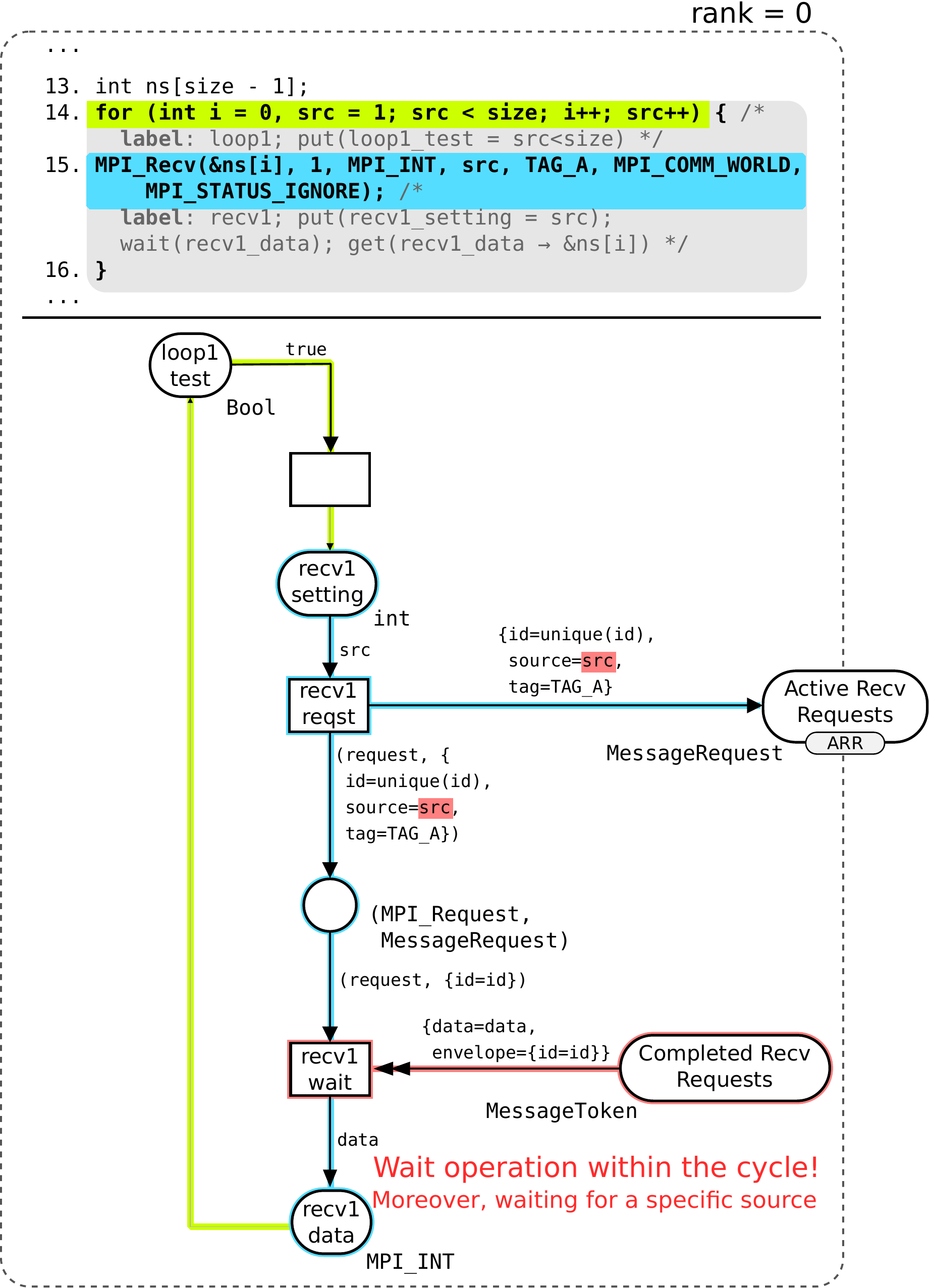}
  }
  \quad
  \subfloat[][]{%
    \label{fig:motiv-example-stage3-rank-zero}
    \includegraphics[width=.44\linewidth]{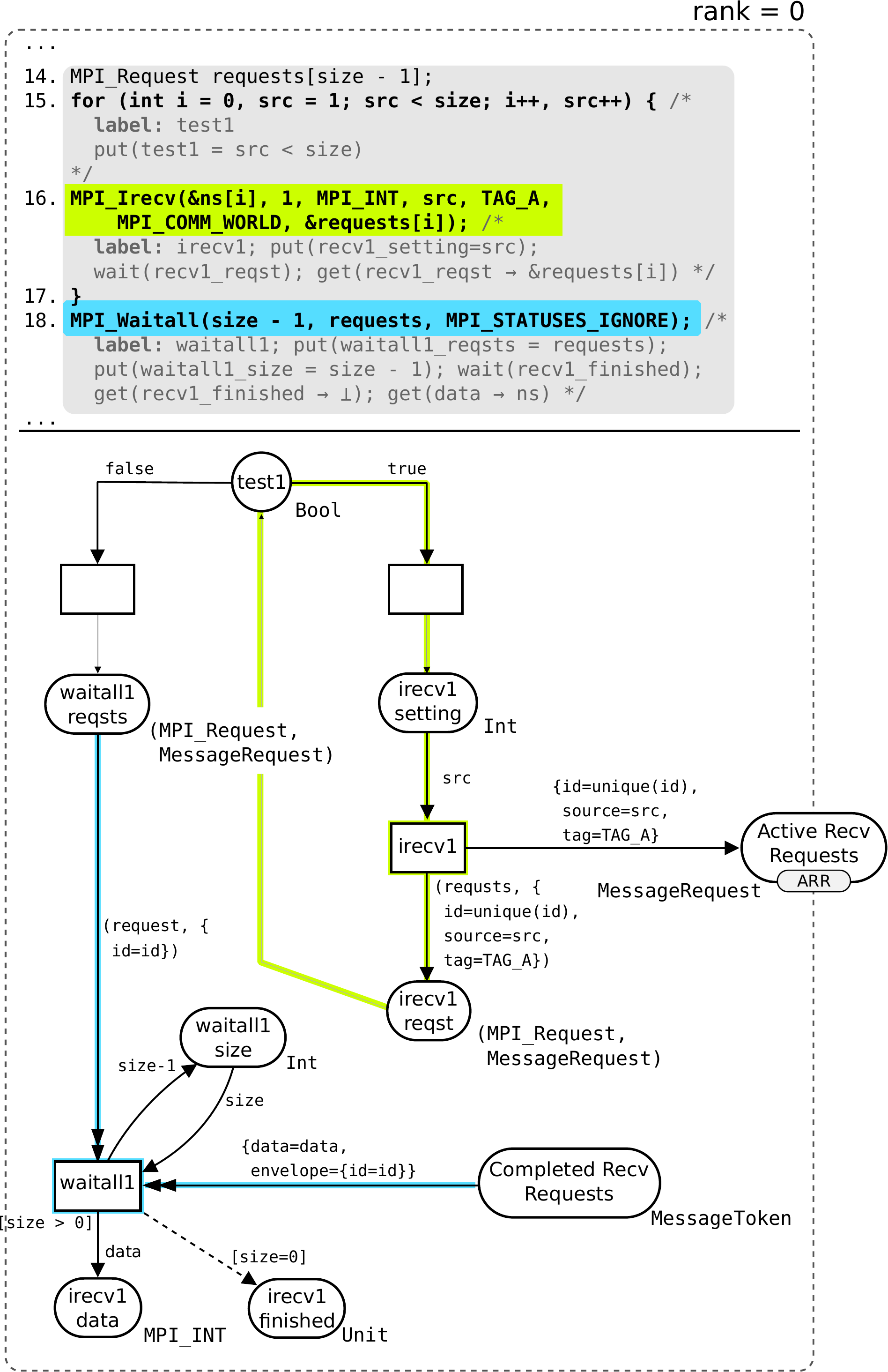}
  }
  \caption[MP net for all-to-one program -- rank zero.]{
    MP net of
    \subref{fig:motiv-example-stage1-rank-zero} the \nth{1} and
    \subref{fig:motiv-example-stage3-rank-zero} \nth{3} version of
    the all-to-one program -- rank zero.
  }
  \label{fig:motiv-example-transformations-rank-zero}
\end{figure}

The second part of MP net (see
\figurename~\ref{fig:motiv-example-stage1-rank-zero}) represents the \nth{0}
process that receives all the messages. It is visible that both sending the
receive request and waiting for its completion are done within the cycle. For
the \nth{2} version (see \lstlistingname~\ref{prg:st-unordered-causal-dependency})
the block of MP net look almost the same, but the \code{source} entry in the
request would be unset, i.e. the request is less restricted.

The second part of MP net (see \figurename~\ref{fig:motiv-example-stage3-rank-zero})
for the \nth{3} version (cf.
\lstlistingname~\ref{prg:st-broken-causal-dependency}) of the program shows that
all the receive requests are sent first (green part) while the waiting for the
completion is done out of the cycle. 

\begin{figure}[ht]
  \centering
  \includegraphics[width=.8\linewidth]{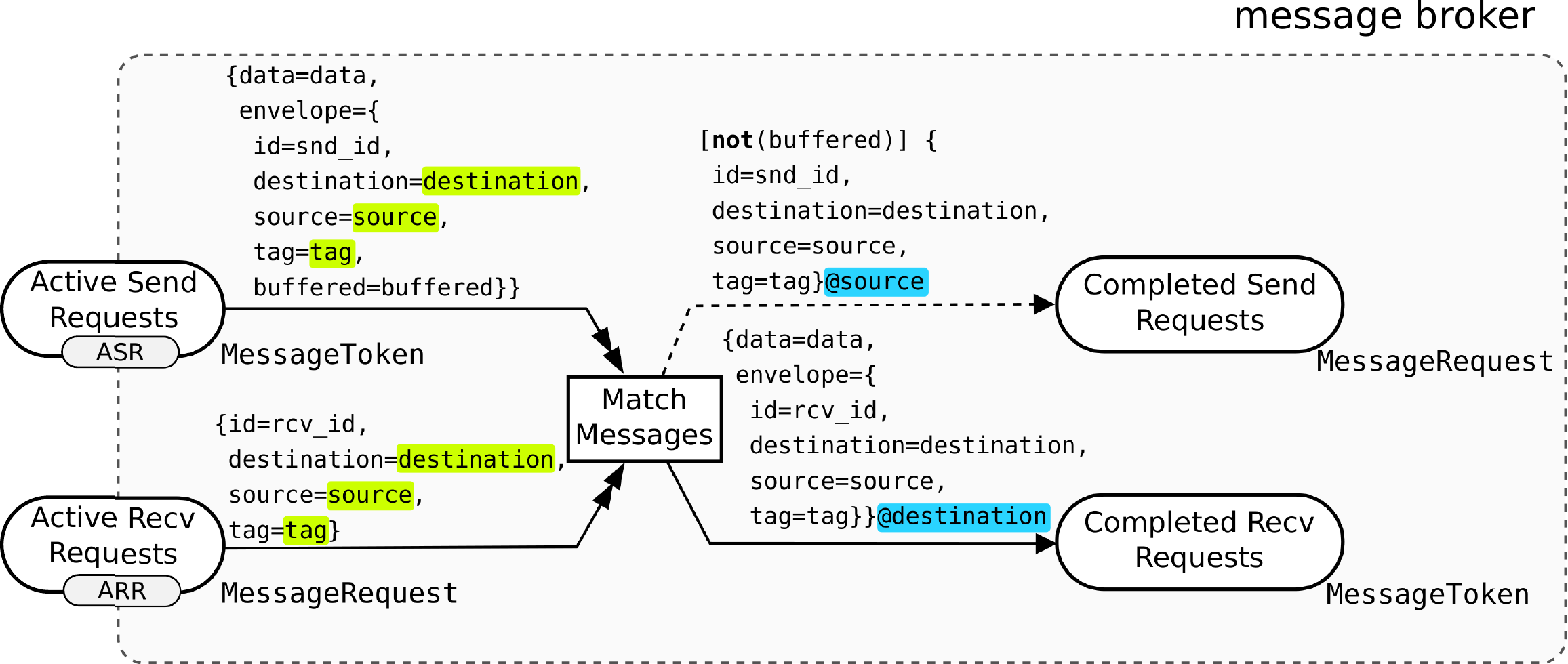}
  \caption{Generic message broker expressed as a part of MP net.}
  \label{fig:message-broker}
\end{figure}

The message broker mentioned at the beginning is also a part of MP net and can
be seen in \figurename~\ref{fig:message-broker}. Nevertheless, it is the same
for all the applications. There is one transition which controls the
\elemname{Active Send/Receive Requests} places and pairs the requests according
to the \code{source}, \code{destination}, and \code{tag} entries. Once these
match, completed requests are generated and sent back (the address specified by
the expression behind `@' sign) to appropriate blocks, into their
\code{Completed Send/Receive Requests} places.

\section{Conclusion}
The main goal of this work is to provide programmers of MPI applications with a
kind of feedback that can help them to quickly check whether or not the actual
communication within their application corresponds to the intended one. For this
purpose MP net is established as a model that should be relatively easily
understood by people and which is capable of capturing the general semantics of
MPI. At the current state, only point-to-point communication is covered.
Nevertheless, the proposed approach does not disable to cover other, more
advanced, routines from MPI. This may be one of a possible future work. However,
the near future is devoted to the implementation of the presented approach into
a working tool.

One could object that the model will increase with the size of the input program.
That is not necessarily true, because the size is directly related to the
number of MPI calls used within the program. Moreover, no matter of the program
complexity MP nets can be still used locally to identify local issues.

Another objection could be that the model is still difficult to read. But the author is
convinced that with a little practice it will be manageable. There is definitely
a learning curve, but in comparison to visual programming tools, users just have to
learn ``to read'' and can benefit from it. Moreover, it may be even more
interesting to use this model further. For example, once it is available,
performance data can be projected into the model, like coloring the most used
transitions or stress the paths that are used most often during the real run. It
can also be used for simulation of the application run and, for example, stop the
program in a situation when two transitions are enabled to see the source of
non-determinism.

As said, the main idea is to provide a model that present the communication in MPI
application in a more comprehensive way. Nevertheless, once the model is
available its potential is much wider than being used only as a static picture
(cf. \figurename~\ref{fig:main-goal}).

\bibliographystyle{acm}
\bibliography{references}

\end {document}